\documentclass[
 reprint,
 amsmath,
 amssymb,
 prb,
 floatfix
]{revtex4-2}

\usepackage{graphicx}
\usepackage{bm}
\usepackage{hyperref}

\renewcommand\Re{\mathop{\rm Re}}

\begin{document}

\title{On chip AC driving for dual Shapiro steps}

\author{David Scheer}
\author{Fabian Hassler}

\affiliation{JARA Institute for Quantum Information, RWTH Aachen University, 52056 Aachen, Germany}

\date{July 2023}

\begin{abstract}
    A single Josephson junction in the phase-slip regime exhibits Bloch oscillations in the voltage when biased with a DC current $I_\text{DC}$. The frequency of the oscillation is given by $\pi I_\text{DC}/e$, with $e$ the elementary charge,  linking the current to the frequency via fundamental constants of nature. If an additional AC drive is applied, the Bloch oscillations may synchronize with the external drive. This leads to the emergence of dual Shapiro steps at fixed current in the $IV$ characteristics of the device. For applications as a current standard, frequencies of the order of 10\,GHz are required. These are challenging to implement experimentally without detrimental effects due to stray capacitances. Here, we propose to employ an additional Josephson junction with a DC voltage bias as an on chip AC source due to the AC Josephson effect. We study the back action of the Bloch oscillations on the Josephson oscillations and identify a parameter regime in which it is minimized. Furthermore, we find that the back action can even be utilized to further enhance the driving signal which can lead to increased widths of the resulting dual Shapiro steps. Finally, we show dual Shapiro steps for a set of realistic experimental parameters at finite temperatures.
\end{abstract}

\maketitle

\section{\label{sec:Introduction}Introduction}

A small Josephson junction in the phase-slip regime~\cite{Mooij2005,Mooij2006,Pop2010,Arutyunov2012,Masluk2012} exhibits Bloch oscillations that relate current to frequency with $I_{\rm DC}=e\omega_B/\pi$~\cite{Likharev1985,Kuzmin1991}. When synchronized to an external AC drive, Bloch oscillations can be used to create dual Shapiro steps in the $IV$ characteristics  at constant current. These steps are of interest in the field of quantum metrology as they provide a current standard only based on frequency and fundamental constants. In this way, they have the potential of closing the last missing link in the quantum metrology triangle~\cite{Averin1985}. 
The voltage standard~\cite{Jeanneret2009} is dual to this as it employs the voltage oscillations created by the AC Josephson effect~\cite{Josephson1962} to create Shapiro steps~\cite{Shapiro1963}. 

Compared to Josephson oscillations, Bloch oscillations are more demanding to  realize experimentally as they suffer from a loss of coherence due to Landau-Zener tunneling~\cite{Geigenmuller1988,Vora2017} and as they need  a high-impedance environment~\cite{Schmid1983,Grabert1992}. The latter can be understood as a consequence of the circuit duality transforming the regime of localized fluxes to a regime of localized charge~\cite{Ulrich2016}. Note that an external AC drive required for dual Shapiro steps is difficult to combine  with the need of a high-impedance environment as  the biasing lines introduce stray capacitances that shunt the phase-slip junction, reducing the impedance of the environment~\cite{ManucharyanPhD,Arndt2018}. 

Attempts to mitigate this problem rely on  on-chip impedances which can be realized either by a large resistance~\cite{Kuzmin1991} or a superinductance~\cite{Corlevi2006,Koch2009,Masluk2012,Marco2015,Arndt2018}. While the former approach introduces additional heating and strong thermal fluctuations, the latter has proven to be viable and enabled first experimental demonstrations of dual Shapiro steps~\cite{Shaikhaidarov2022,Crescini2023}. 

In this paper, we present an approach to produce dual Shapiro steps that does not rely on an external AC drive. We propose a circuit, first analyzed in \cite{Hriscu2013}, that is able to generate the required AC driving voltage on chip by using the AC Josephson effect. The approach is similar to the synchronization of Bloch oscillations~\cite{Kaap2023}. However, an important difference  is the fact that the coherence of Josephson oscillations is not limited by Landau-Zener tunneling. This allows to use the Josephson oscillations as a more rigid drive signal instead of relying on a mutual synchronization with a small DC bias on both junctions. We study the effect the back action of the Bloch oscillations on the Josephson oscillations and identify the parameter regime where the back action is negligible. We investigate the effects of a finite back action numerically. We show that the back action can increase the width of the dual Shapiro steps. We identify realistic circuit parameters for the observation of dual Shapiro steps and show their robustness towards thermal noise. 

\begin{figure}[tb]
\includegraphics{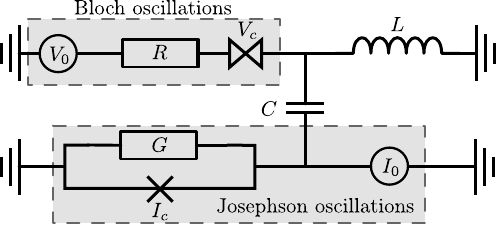}
\caption{\label{fig:setup} Circuit consisting of an overdamped phase-slip junction (gray box in the upper branch) and an overdamped Josephson junction (gray box in the lower branch). These elements are connected by an $LC$-resonator acting as a transconductance.  Note that both the  current $I_0$ and the  voltage $V_0$ are solely DC biased.}
\end{figure}

\section{Circuit setup}

The simplest circuit to realize Bloch oscillations is a phase-slip junction connected in series with a resistance $R$ and  a DC bias voltage $V_0$, see Fig.~\ref{fig:setup}.
In the quasiclassical regime, with $R\gg R_{\rm Q}=h/4e^2=6.45\,{\rm k}\Omega$, the quantum fluctuations of the charge operator $\hat Q$ in the loop are below the charge $2e$ of a  Cooper pair. Therefore, the Heisenberg equation of the charge is  well approximated by the equation of motion
\begin{equation}
\label{eq:Bloch_circuit}
V_0=R\dot Q+V_c\sin(\pi Q/e)
\end{equation}
of the expectation value $Q=\langle \hat Q\rangle$, where $V_c$ is the critical voltage of the phase-slip junction \cite{Arndt2018}. 
The characteristic rate $\omega_R=\pi V_c/eR$ of the equation  corresponds the $RC$-time of the linearized capacitance of the phase-slip junction. For a given DC bias $V_0$, the frequency of the Bloch oscillations $\omega_B$ is given by
\begin{equation}
   \omega_B= \omega_R\sqrt{(V_0/V_c)^2-1}\,.
\end{equation} 
The dual circuit that realizes Josephson oscillations is also shown in Fig.~\ref{fig:setup}. For a given bias current $I_0$, its  equation of motion for the node flux $\Phi$
\begin{equation}
    I_0=G\dot\Phi+I_c\sin (\pi \Phi/e R_{\rm Q})\,,
\end{equation}
is dual to  Eq.~\eqref{eq:Bloch_circuit} with a critical current $I_c$ and a conductance $G\gg R_{\rm Q}^{-1}$.
The relevant rate  $\omega_G=2eI_c/\hbar G$  corresponds to an $RL$-time for a linearized Josephson junction. The frequency of the dual Josephson oscillations is determined by the bias current $I_0$ according to
\begin{equation}
    \omega_J=\omega_G\sqrt{(I_0/I_c)^2-1}\,.
\end{equation}
We are interested in the synchronization of these oscillations that appears when their frequencies match with $\omega_B = \omega_J$.
The ratio  $\omega_R/\omega_G$  thus determines the required biases $V_0, I_0$.
The regime where a rigid Josephson oscillation in the lower branch drives the Bloch oscillations in the upper branch demands 
$\omega_R\gg \omega_G$ such that  $I_0 \gg I_c$ while $V_0 \gtrsim V_c$.  For a Josephson junction in the phase-slip regime, the latter condition assures that Landau-Zener tunneling into higher energy bands is suppressed \cite{Geigenmuller1988,Arndt2018}.

The two elementary circuits are coupled via an $LC$-resonator that acts as a transconductance, see Fig.~\ref{fig:setup}. This circuit has been analyzed in the context of the mutual synchronization of two conjugate quantum variables~\cite{Hriscu2013}. Here, we investigate the regime where we the Josephson oscillations take the role of a rigid AC driving signal.  The coupled circuits in the overdamped regime obey a set of equations of motion
\begin{align}I_0&\!=\!G\dot \Phi+I_c\sin(\pi \Phi/e R_{\rm Q})+C\dot V_{C},
& \dot Q&\!=\!I_L+C\dot V_{C}, \nonumber
\\
V_0&\!=\!R\dot Q+V_c\sin(\pi Q/e)
+L\dot I_L,
&\dot\Phi&\!=\!V_{C}-L\dot I_L,
\label{eq:not_self_dual}
\end{align}
where $I_L$ is the current through the inductance $L$ and $V_C$ is the voltage across the capacitance $C$. Note that even though it connects two circuits that are dual to each other by a self dual element, the circuit is not self dual which is reflected in the negative sign in the last equation of Eqs.~\eqref{eq:not_self_dual}.

The relevant scales for current and voltage are set by the critical voltage $V_c$ and the critical current $I_c$ which give rise to a natural resistance scale $R_c=V_c/I_c$. 
We describe the effect of the $LC$ oscillator by the characteristic impedance $Z_0=\sqrt{L/C}$ and the resonance frequency $\omega_0=1/\sqrt{LC}$. In addition to the resonance frequency, the $LC$-resonator introduces  new relaxation timescales for the Bloch oscillation: an $RL$-time $\tau_L=L/R$ in the upper branch and a $RC$-time $\tau_C=C/G$ from the connection to the lower branch.

In order to produce dual Shapiro steps, the AC Josephson oscillations should function as a rigid drive signal. For this the displacement current $\simeq C \dot V_C$ that is coupled out of the lower branch should be small compared to bias current $I_0$, \emph{i.e.}, $C\dot V_{C}\ll I_0$. In this regime, the dominant contribution of the phase $\varphi=\pi\Phi/e R_{\rm Q}$ increases linearly in time with $\varphi=(I_0/I_c)\omega_G t+\varphi_1$, where $\varphi_1$ is a small correction  $\dot\varphi_1\ll\omega_GI_0/I_c$. As a result, the equation for the Josephson circuit simplifies to
\begin{equation}
    \sin{\omega_J t}+\frac{\dot\varphi_1}{\omega_G}+\frac{C\dot V_{C}}{I_c}=0\,,
\end{equation}
with $\omega_J\approx \omega_G I_0/I_c$. 

Using the Fourier transformed equations of motion, we are able to integrate the other degrees of freedom of the circuit to derive an effective equation of motion for the dimensionless charge $q=\pi Q/e$  on the phase-slip junction.
If the frequency of the Josephson oscillations matches the resonance frequency of the $LC$-resonator ${\omega_J=\omega_0}$, the effective equation of motion is given by
\begin{equation}
    \label{eq:effective_model}
    \frac{V_0}{V_c}-\frac{Z_0}{R_c}\cos{\omega_0 t}=\frac{\dot{q}}{\omega_R}+\sin{q}+F(q,t)
\end{equation}
with a term 
\begin{equation}
\label{eq:back_action}
F(q,t)=\frac{\tau_L}{\omega_R}\int\frac{d\omega}{2\pi}\frac{1-i\omega\tau_C}{1-i\omega\tau_C-\omega^2/\omega_0^2} (\ddot q)_\omega e^{-i\omega t}\,,
\end{equation}
that describes the modification of the drive by the back action from the Bloch oscillations onto the Josephson oscillations with  $ (\ddot q)_\omega=\int \ddot q(t)e^{i\omega t}dt\,$.
Up to the back action term, the resulting equation corresponds to that of an overdamped phase-slip junction with a DC bias and an AC driving voltage of strength $Z_0/R_c$.  Note that the resonance condition is only used to obtain the maximal driving strength and has no effect on the form of the back action term.

As shown in App.~\ref{appa}, the back action $F(q,t)$ is suppressed in the regime $\omega_0\tau_L\ll1$ and $\tau_L\ll\tau_C$. The first condition, equivalent to $Z_0\ll R$, can be understood by thinking about  the current divider in the upper branch; it ensures that the current propagating from the Josephson part of the circuit mainly goes through the inductance realizing an effective AC voltage in the upper loop. The second condition $\tau_L\ll\tau_C$ compares the relaxation timescales of currents remaining in the upper branch and currents going into the lower branch. If the upper branch is much faster to adapt to changes than the lower branch, the dynamics of the Josephson oscillations essentially decouple from the dynamics of the Bloch oscillations rendering them a stable AC voltage source.

In the regime of large driving frequencies $\omega_J= \omega_0\gg\omega_R$ and small effective driving strengths $Z_0/R_c\ll1$ the dual Shapiro steps with suppressed back action have a width $\Delta V$ of
\begin{equation}\label{eq:step_width}
    \frac{\Delta V}{V_c}=\frac{\omega_R Z_0}{\omega_0 R_c}+\mathcal{O}(\omega_R^2/\omega_0^2),
\end{equation}
as we derive in App.~\ref{appa}.
This result is in agreement with the result for a rigid external drive of strength $Z_0/R_c$~\cite{Likharev1986}. 
\section{Simulation of dual Shapiro steps}

Due to the nonlinearities in Eqs.~\eqref{eq:not_self_dual}, it is difficult to gain analytical insights into the width of dual the Shapiro steps that go beyond perturbation theory for small driving strengths given above. Since we expect the largest  steps outside this regime, it is crucial to perform a numerical analysis of the problem to find optimal parameters for an experimental implementation of our proposal.

\begin{figure}[tb]
\includegraphics{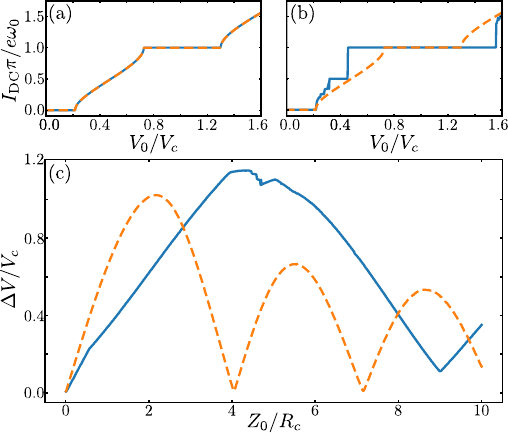}
\caption{\label{fig:ba_comparison} Width of the dual Shapiro steps for the full model (solid) and for the reduced model without back action (dashed)  for $\omega_0/\bar\omega=10$, $\omega_R/\omega_G=100$, and $\omega_0\tau_C=10$. We show the numerical results for the first dual Shapiro step for $Z_0/R_c=5$ with $\omega_0\tau_L=0.01$ in  (a) and   $\omega_0\tau_L=2$  in (b). For $\omega_0\tau_L=0.01$ no effect of the back action is visible whereas the back action is important for $\omega_0\tau_L=2$. We investigate the effect of the back action ($\omega_0\tau_L=2$) further in (c), where we provide the width of the first dual Shapiro step for different effective driving strengths $Z_0/R_c$. We see that the back action increases the width for $Z_0/R_c$ between 3 and 8. Moreover, in particular around $Z_0/R_c =4$, the back action leads to multiple kinks that originate from a competing synchronization step at half  the target frequency.}
\end{figure}

Measuring the currents (voltages) in units of $I_c$ ($V_c$), our circuit in the quasiclassical regime is described by the five dimensionless parameters  $Z_0/R_c$, $\omega_0/\bar\omega$,  $\omega_0\tau_L$, $\omega_0\tau_C$, and $\omega_R/\omega_G$, 
where we introduced the rate $\bar\omega=\sqrt{\omega_R\omega_G}$, given by the geometric mean of the characteristic rates in the two sub-circuits.  In order to simulate the $IV$ characteristic of the upper branch, we set the bias current (producing an effective AC drive) in the lower branch to $I_0=I_c\omega_0/\omega_G$. This choice of bias current assures that the Josephson oscillations are on resonance with $\omega_J \approx (I_0/I_c)\omega_G = \omega_0$.
We record the DC current $I_\text{DC} = \int_0^\tau \dot Q dt / \tau$, $\tau \to \infty$, across the upper branch for a sweep of the bias voltage $V_0$. We integrate the equations of motion Eqs.~\eqref{eq:not_self_dual} using the forward Euler method with discrete time steps of size $\delta=5\times10^{-4}/\bar\omega$ in order to resolve all the relevant timescales of the circuit.

In Fig.~\ref{fig:ba_comparison}, we compare the dual Shapiro steps arising in the full circuit with steps resulting from the effective model in Eq.~\eqref{eq:effective_model} without the back action term, \emph{i.e.}, with $F \equiv 0$, for the parameters $\omega_0/\bar\omega=10$, $\omega_R/\omega_G=100$, and $\omega_0\tau_C=10$. The $IV$ characteristic for $Z_0/R_c=5$ with $\omega_0\tau_L=0.01$ is depicted in Fig.~\ref{fig:ba_comparison}(a) where the effect of back action is small. In Fig.~\ref{fig:ba_comparison}(b), we show that for $\omega_0\tau_L=2$  the back action is relevant.
In particular, with the back action, the dual Shapiro step is wider and pronounced sub-harmonic steps occur that are discussed in details in  \cite{Hriscu2013}.
In order to investigate the effect of the back action in more details, Fig.~\ref{fig:ba_comparison}(c) shows the comparison of the step width $\Delta V$ with and without the back action for varying $Z_0/R_c$.
In particular, we see that for $Z_0/R_c$ between 3 and 8, the back action is favorable and increases the width of the first dual Shapiro step.
\begin{figure}[tb]
\includegraphics{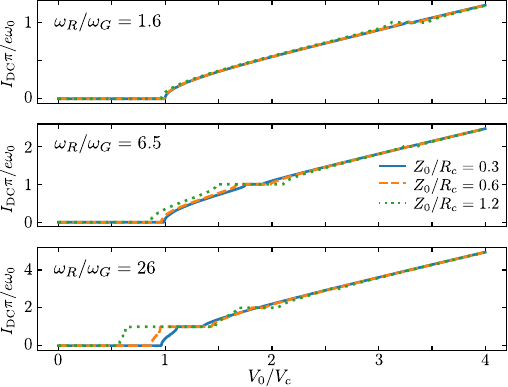}
\caption{\label{fig:shifting_steps} Dual Shapiro steps at $\omega_0/\bar\omega=4$, $\omega_0\tau_C=0.04$, and $\omega_0\tau_L=0.004$ for varying $Z_0/R_c$ and $\omega_R/\omega_G$. For increasing $\omega_R/\omega_G$, the position of the step is moved closer to the region of the Coulomb blockade. This increases the sensitivity of the steps to the effective driving strength $Z_0/R_c$ and decreases the bias voltage required to create the steps.}
\end{figure}
Note, however, that  presently it appears experimentally rather difficult to realize circuits with $\omega_0\tau_L\gtrsim 1$ that feature an appreciable effect due to the back action while still remaining in the quasiclassical regime. The reason is that quasiclassics demands that $R \gg R_Q $ while $\omega_0\tau_L\gtrsim 1$ corresponds to $Z_0 \gtrsim R$. Presently, $Z_0$ is at most of order k$\Omega$ such that the regime with enhanced back action cannot be realized for any $R$. 

\begin{figure}[tb]
\includegraphics{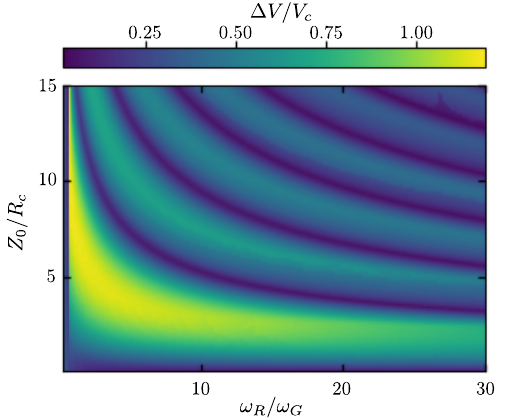}
\caption{\label{fig:step_widths} Width of the first dual Shapiro step for varying $Z_0/R_c$ and $\omega_R/\omega_G$ with the  parameters of Fig.~\ref{fig:shifting_steps}. At constant $\omega_R/\omega_G$, the step width oscillates in the effective driving strength $Z_0/R_c$ with faster oscillations for increasing $\omega_R/\omega_G$. While the slow oscillations at small $\omega_R/\omega_G$ allow for more stable steps with regards to an imprecise impedance, the impedance required for a visible step is also increased.}
\end{figure}

To understand the effects of the effective driving strength $Z_0/R_c$ and the ratio of the characteristic rates $\omega_R/\omega_G$, we investigate the behavior of the circuit for parameters $\omega_0/\omega_R=4$, $\omega_0\tau_C=0.04$, and $\omega_0\tau_L=0.004$. In Fig.~\ref{fig:shifting_steps}, the $IV$ characteristics for different values of $Z_0/R_c$ and $\omega_R/\omega_G$ are depicted. The curves exhibit an increased step width at a lower voltage bias with increasing $Z_0/R_c$ and $\omega_R/\omega_G$. 
In Fig.~\ref{fig:step_widths}, we show the resulting step widths as a function of $Z_0/R_c$ and $\omega_R/\omega_G$. At a fixed ratio $\omega_R/\omega_G$, the step widths exhibit oscillations in the effective driving strength $Z_0/R_c$ similar to Fig.~\ref{fig:ba_comparison}(c). Note that the period of oscillation decreases  for increasing $\omega_R/\omega_G$. The reason for this is that at $\omega_R\gg\omega_G$, the dual Shapiro steps are more sensitive to the amplitude of the effective driving as they appear closer to the Coulomb blockade region of the phase-slip junction where the nonlinearity is important.

\section{Discussion of realistic experimental parameters}

When trying to observe dual Shapiro steps in an experimental setup, there are two major effects that lead to incoherent Bloch oscillations and therefore impair their ability for synchronization. The first is the occurrence of Landau-Zener tunneling into higher bands of the Josephson junction in the phase-slip regime that invalidates the model of adiabatic ground state transport used for the description as a circuit element. When the current through the junction becomes too large, these processes can no longer be neglected. For a typical junction with  $V_c=50\,\mu$V, Landau-Zener tunneling can be neglected at a current much smaller than 20\,nA \cite{Geigenmuller1988}. Therefore, the resonance frequency of the oscillator should fulfill $e\omega_0/\pi \ll 20$\,nA.
The second effect are the thermal fluctuations that lead to Nyquist-Johnson noise originating from the biasing resistors which limits the stability of the voltage and current bias. In our simulations, we implement the thermal fluctuations at a temperature $T$ as Gaussian white noise increments with strength $2Rk_BT/\sqrt{\delta}$ ($2Gk_BT/\sqrt{\delta}$) added to the bias voltage (current).

\begin{figure}[tb]
\includegraphics{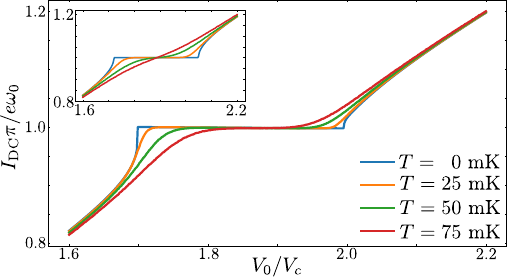}
\caption{First dual Shapiro step for realistic circuit parameters given by $L=0.01\,\mu\rm H$, $C=100\,\rm fF$, $R=50\,{\rm k}\Omega$, $G=0.1\,\rm S$, $I_c=0.1\,\mu \rm A$, and $V_c=50 \,\mu\rm V$ with a resulting on-step current of 1.6\,nA. The dimensionless parameters are the same as in Fig.~\ref{fig:shifting_steps} with  $\omega_R/\omega_G = 6.5$ and $Z_0/R_c = 0.6$ (corresponding to the dashed orange line in the center panel). Note that even though the back action is small it renders the dual Shapiro step is more robust against thermal noise compared to the case without back action (inset).}\label{fig:experimental_params}
\end{figure}

Figure~\ref{fig:experimental_params} shows dual Shapiro steps at varying temperatures for experimentally realistic circuit parameters given by $L=0.01\,\mu$H, $C=100\,$fF, $R=50\,$k$\Omega$, $G=0.1\,$S, $I_c=0.1\,\mu$A, and $V_c=50\,\mu$V \cite{sergey_fabian}. These circuit parameters correspond to $Z_0/R_c\approx 0.6$, $\omega_0/\bar\omega\approx 4$,  $\omega_0\tau_C\approx 0.04$, $\omega_0\tau_L\approx 0.004$, and $\omega_R/\omega_G \approx 6.5$, cf.\ Fig.~\ref{fig:shifting_steps}. In this regime, the effect of back action is small and the step height is approximately given by $\Delta V/V_c \approx \omega_R Z_0/\omega_0 R_c \approx  0.4 $, see Eq.~\eqref{eq:step_width}. The simulation shows a clearly visible first dual Shapiro step that gets smeared out at finite temperature.
The resulting on-step current is given by $1.6\,$nA which indicates that Landau-Zener tunneling is not yet a relevant factor.

\section{Conclusion}

In conclusion, we have analyzed the emergence of dual Shapiro steps in a circuit that uses Josephson oscillations to drive a phase-slip junction via an transconductance build by an $LC$-resonator. Our setup only requires DC wiring to produce dual Shapiro steps. We described the system by a set of coupled equations of motion neglecting the influence of the higher bands of the Josephson junction in the phase-slip regime. In the regime of a large current bias $I_0\gg I_c$, we derived an effective model for the charge dynamics of the phase-slip junction. Up to a back action term, this model is equivalent to an overdamped phase-slip junction with an external AC drive of strength $Z_0/R_c$. In the experimentally relevant regime of $Z_0\ll R$ and $\tau_L\ll\tau_C$, the back action is suppressed. As a result, our setup realizes a rigid AC drive without the need for external AC biasing lines.  A numerical analysis of the first dual Shapiro step outside this regime shows that a finite back action can lead to enhanced step widths as well additional sub-harmonic synchronization steps. Finally, we presented a set of realistic experimental parameters that yield a dual Shapiro step at 1.6\,nA which is robust to thermal Nyquist-Johnson noise at temperatures up to 75\,mK. Throughout this work, we have neglected the effects of Landau-Zener tunneling which is valid for small currents.  A detailed analysis of these effects which become important for elevated currents is an important question for future research.

\begin{acknowledgments}
We thankfully acknowledge fruitful discussions with S.~Lotkhov and F.~Kaap.
This work was supported by the Deutsche Forschungsgemeinschaft (DFG) under
Grant No. HA 7084/6--1.
\end{acknowledgments}

\appendix

\section{Realizing an effective AC drive}
\label{appa}
In this appendix, we examine the dynamics of the charge in the phase-slip junction described by Eq.~\eqref{eq:effective_model}. We consider the regime of large driving frequencies $\omega_J= \omega_0\gg\omega_R$, where we analyze the emergence of dual Shapiro steps with the added effects of back action on the driver described by Eq.~\eqref{eq:back_action}.
We solve the equation of motion by making an ansatz $q(t)=q_0(t)+q_1(t)$ based around Bloch oscillations at the frequency $\omega_0$ without back action or AC drive~\cite{Likharev1986}
\begin{equation}
q_0(t)=2 \arctan\Biggl\{\frac{\omega_0}{\sqrt{\omega_0^2+\omega_R^2}-\omega_R}\tan\biggl[\frac{1}{2}(\omega_0 t+\theta_0)\biggr]\Biggr\}\,.
\end{equation}
We allow $\theta_0$ to be time dependent but assume that it is slow on the scale of $\omega_0$. We add a small perturbation
\begin{equation}
    q_1(t)=A_1 \Re{\left[e^{-i(\omega_0 t+\theta_1)}\right]},
\end{equation}
with amplitude $A_1>0$ and phase $\theta_1$
as a linear response to the driver with $A_1\ll 1$.
This ansatz allows us to separate Eq.~\eqref{eq:effective_model} into a couple of equations
\begin{align}
 \frac{\tilde V_0}{V_c}&=\frac{1}{\omega_R}\dot q_0+\sin q_0 ,
 \label{eq:unperturbed}\\
 \frac{\Delta\tilde V_0}{V_c}&=\frac{1}{\omega_R}\left(\frac{\dot \theta_0}{\omega_0}\dot q_0+\dot q_1\right)  +q_1\cos q_0 +\frac{Z_0}{R_c}\cos\omega_0 t \nonumber\\
 &\quad+F(q_0,t),
 \label{eq:perturbation}
\end{align}
where we split the bias voltage $V_0$  into two parts, $V_0 = \tilde V_0 + \Delta \tilde V_0$. The main part $\tilde V_0=V_c\sqrt{\omega_0^2/\omega_R^2+1}$ leads to Bloch oscillations at frequency $\omega_0$ as the solution to Eq.~\eqref{eq:unperturbed}. The deviation $\Delta\tilde V_0$ from this DC bias must be compensated by the remaining terms collected in Eq.~\eqref{eq:perturbation} which are small enough to still preserve the Bloch oscillations. In the case of slow $\theta_0$, the main DC contribution originates from the term $q_1\cos q_0$ which allows a down-conversion of the linear response to a DC voltage.

In order for the charge dynamics in the circuit to match the dynamics of an overdamped phase-slip junction with a rigid AC drive, the back action term $F(q_0,t)$ must only give a small contribution. The back action term can only contain frequencies that are already present in $\ddot q_0$ which are the integer multiples of $\omega_0$, where the connecting $LC$-resonator suppresses the off-resonant higher order contributions. In the regime $\omega_0\gg\omega_R$, the contributions at $k\omega_0$ are further suppressed by $(\omega_R/\omega_0)^k$ in $\ddot q_0$ itself.
The resulting back action is is given by
\begin{equation}
    F(q_0,t)=-\frac{\tau_L}{\tau_C}\Bigl\{\Re\Bigl[(\omega_0\tau_C-i)e^{-i(\omega_0 t+\theta_0)}\Bigr]+\mathcal{O}(\omega_R/\omega_0)\Bigr\},
\end{equation}
which is a small contribution in the regime of $\omega_0\tau_L\ll1$ and $\tau_L\ll\tau_C$. The back action modifies the AC drive on the system with the dominant correction in phase with the driving term. This yields a modified driving amplitude of
\begin{equation}
    \frac{\tilde V(\theta_0)}{V_c}=\frac{Z_0}{R_c}-\frac{\tau_L}{\tau_C}(\omega_0\tau_C\cos\theta_0+\sin\theta_0),
\end{equation}
that is dependent on the phase of the Bloch oscillations. 

Next, we average Eq.~\eqref{eq:perturbation} for frequencies close to $0$ and $\omega_0$. We   obtain the conditions
\begin{align}
    \frac{\Delta\tilde V_0}{V_c}&=\frac{\dot\theta_0}{\omega_R}+\frac{A_1}{2}\cos(\theta_0-\theta_1)\,,\\
    A_1e^{-i(\theta_1-\pi/2)}&=\frac{\omega_R}{\omega_0}\left[\frac{Z_0}{R_c}-\frac{\tau_L}{\tau_C}(\omega_0\tau_C\cos\theta_0+\sin\theta_0)\right],\nonumber
\end{align}
up to terms of $\mathcal{O}(\omega_R^2/\omega_0^2)$.
To calculate the width of the first dual Shapiro step, we set $\dot\theta_0=0$ which corresponds to a synchronization with the drive. The step corresponds to the region
\begin{align}
    -\frac{\omega_R}{2\omega_0}\left(\frac{Z_0}{R_c}+\frac{\tau_L}{\tau_C}\right)\leq \frac{\Delta\tilde V_0}{V_c} \leq \frac{\omega_R}{2\omega_0}\left(\frac{Z_0}{R_c}-\frac{\tau_L}{\tau_C}\right),
\end{align}
where the change in the DC bias can be compensated by a phase locking between the drive and the Bloch oscillations which yields a step of width 
\begin{equation}
    \frac{\Delta V}{V_c}=\frac{\omega_R Z_0}{\omega_0 R_c}+\mathcal{O}(\omega_R^2/\omega_0^2),
\end{equation}
as in the case of a rigid AC drive~\cite{Likharev1986}.
While the step width remains unchanged by the back action, the position of the step is shifted by
\begin{equation}
    \frac{\delta V}{V_c}=-\frac{\omega_R \tau_L}{2\omega_0 \tau_C}+\mathcal{O}(\omega_R^2/\omega_0^2).
\end{equation}
Overall, the circuit realizes the model of a rigid AC drive with a small correction to the position of the first dual Shapiro step in the regime of large driving frequencies and a short $RL$ timescale $\tau_L$ in the upper branch of the circuit.

\end{document}